\documentclass[conference]{IEEEtran}
\IEEEoverridecommandlockouts
\usepackage{cite}
\usepackage{amsmath,amssymb,amsfonts}
\usepackage{algorithmic}
\usepackage{graphicx}
\usepackage{textcomp}
\usepackage{xcolor}
\def\BibTeX{{\rm B\kern-.05em{\sc i\kern-.025em b}\kern-.08em
    T\kern-.1667em\lower.7ex\hbox{E}\kern-.125emX}}

\begin{document}

\title{Segmenting Medical Images: From UNet to Res-UNet and nnUNet}

\author{\IEEEauthorblockN{Lina Huang, Alina Miron, Kate Hone and Yongmin Li}
\IEEEauthorblockA{\textit{Deptment of Computer Science},
\textit{Brunel University London},
Uxbridge, UK \\
\{lina.huang, alina.miron, kate.hone, yongmin.li\}@brunel.ac.uk}
}

\maketitle
\begin{abstract}
This study provides a comparative analysis of deep learning models—UNet, Res-UNet, Attention Res-UNet, and nnUNet—evaluating their performance in brain tumour, polyp, and multi-class heart segmentation tasks. The analysis focuses on precision, accuracy, recall, Dice Similarity Coefficient (DSC), and Intersection over Union (IoU) to assess their clinical applicability. In brain tumour segmentation, Res-UNet and nnUNet significantly outperformed UNet, with Res-UNet leading in DSC and IoU scores, indicating superior accuracy in tumour delineation. Meanwhile, nnUNet excelled in recall and accuracy, which are crucial for reliable tumour detection in clinical diagnosis and planning. In polyp detection, nnUNet was the most effective, achieving the highest metrics across all categories and proving itself as a reliable diagnostic tool in endoscopy. In the complex task of heart segmentation, Res-UNet and Attention Res-UNet were outstanding in delineating the left ventricle, with Res-UNet also leading in right ventricle segmentation. nnUNet was unmatched in myocardium segmentation, achieving top scores in precision, recall, DSC, and IoU. The conclusion notes that although Res-UNet occasionally outperforms nnUNet in specific metrics, the differences are quite small. Moreover, nnUNet consistently shows superior overall performance across the experiments. Particularly noted for its high recall and accuracy, which are crucial in clinical settings to minimize misdiagnosis and ensure timely treatment, nnUNet's robust performance in crucial metrics across all tested categories establishes it as the most effective model for these varied and complex segmentation tasks.
\end{abstract}
\section{Introduction}
Medical image segmentation, the process of partitioning an image into multiple segments to identify regions of interest, plays a crucial role in various clinical applications, including disease diagnosis, treatment planning, and computer-assisted surgery \cite{b1, b2, b3, b4}. However, accurate and robust segmentation is a challenging task due to the complexity and variability of medical images, often complicated by factors such as noise, intensity inhomogeneity, and low contrast \cite{b5, b6}. Traditional segmentation techniques, such as thresholding, region growing, and active contours, have limitations in handling these challenges \cite{b7}, motivating the development of advanced methods. In recent years, deep learning techniques have emerged as powerful tools for medical image analysis, offering the potential to address these challenges effectively \cite{b8}.

Deep learning, particularly convolutional neural networks (CNNs), has emerged as a powerful tool for medical image segmentation, demonstrating superior performance over conventional approaches\cite{b4, b9, b10}. Among the various CNN architectures, the UNet \cite{b11} and its variants, like Res-UNet \cite{b12}, Attention Res-UNet \cite{b13} and nnUNet \cite{b14} have gained significant attention due to their ability to capture multi-scale features and effectively segment complex structures. 

While these models have demonstrated promising results in various applications, their relative performance on specific medical image segmentation tasks remains an area of active research. Evaluating and comparing these approaches is crucial for selecting the most suitable model for a given task and identifying potential areas for further improvement.

This study aims to comprehensively evaluate the performance of the UNet, Res-UNet, Attention Res-UNet, and nnUNet models on three medical image segmentation tasks: brain tumour segmentation, polyp segmentation, and cardiac segmentation. Through comparative analysis, this research seeks to identify the strengths and weaknesses of each model, determining which performs best across various segmentation scenarios. Such insights aim to set a benchmark in the field of medical image segmentation and provide practical guidance to future researchers on key considerations when applying UNet and its variants to medical image analysis. 

The paper is structured as follows: Section 2 reviews related work in medical image segmentation and deep learning. Section 3 outlines experimental setup on  brain tumour, colorectal polyp, and cardiac segmentation. Sections 4 to 6 present results. Section 7 provides the discussion, and Section 8 presents the conclusions of this study.

\section{Background}
With the advent of advanced medical imaging modalities, such as X-ray (including mammography), computed tomography (CT), magnetic resonance imaging (MRI), positron emission tomography (PET), and ultrasound, physicians have access to highly detailed and informative visual representations of the human body's internal structures and physiological processes.
Nonetheless, the pathway to becoming a radiologist involves years of specialized training and extensive clinical practice, leading to a notable shortage of professionals in this field \cite{b15, b16}. This upsurge in demand has consequentially amplified the workload for radiologists, potentially heightening the incidence of diagnostic errors and misdiagnoses due to fatigue and burnout \cite{b15, b16, b17}.\\

Alternatively, the endeavor of accurately and reliably segmenting regions of interest from these varied imaging modalities presents considerable challenges. These challenges stem from the inherent complexity and variability of medical images, further complicated by factors like noise, uneven intensity distributions, low contrast, and the presence of artifacts \cite{b5, b6}. For instance, X-ray and CT images may suffer from beam hardening and scatter artifacts, while MRI images can be affected by intensity inhomogeneities and geometric distortions. Ultrasound images, on the other hand, are frequently plagued by speckle noise and shadowing artifacts, rendering segmentation a daunting task.

Historically, segmentation techniques such as thresholding, region growing, and active contours have seen considerable application \cite{b17}. While these methods have demonstrated utility in certain scenarios, they often falter when dealing with the intricacies of complex medical images, particularly in the presence of noise, intensity variations, and ambiguous boundaries \cite{b7}. Moreover, these approaches typically rely on hand-crafted features and prior knowledge, limiting their generalizability and adaptability to diverse imaging modalities and anatomical structures\cite{b7}. Statistical methods have also been extensively applied to medical image segmentation, including probabilistic modelling \cite{kaba2014retinal,kaba2013segmentation}, Markov Random Field \cite{salazar2012mrf}, graph-cut \cite{salazar2014segmentation,salazar2011optic,salazar2010retinal,dodo2018graph}, level set \cite{wang2015level,dodo2019retinal}. 


In recent years, deep learning techniques, particularly convolutional neural networks (CNNs), have revolutionized the field of medical image analysis, offering powerful solutions for segmentation tasks across various imaging modalities \cite{b1, b18}. CNNs possess the remarkable ability to automatically learn hierarchical feature representations from raw image data, enabling accurate and robust segmentation of complex anatomical structures \cite{b19, b20}. This data-driven approach alleviates the need for hand-crafted features and prior knowledge, making it more generalizable and adaptable to various imaging modalities and anatomical structures. Automated segmentation methods based on deep learning can alleviate the workload on radiologists, improve efficiency, and reduce the risk of misdiagnosis, ultimately enhancing the clinical workflow and patient care. Among the various CNN architectures, the UNet \cite{b11} and its variants, like Res-UNet \cite{b12}, Attention Res-UNet \cite{b13} and nnUNet \cite{b14} have gained significant attention due to their ability to capture multi-scale features and effectively segment complex structures.

Despite the remarkable progress made by deep learning models in medical image segmentation, several challenges persist. Existing models may struggle with accurately segmenting small or irregularly shaped lesions, preserving fine details, and maintaining robustness across diverse imaging modalities and clinical scenarios. Additionally, the successful deployment of these models often requires substantial computational resources, large-scale annotated datasets, and extensive manual tuning of hyperparameters, hindering their widespread adoption in clinical settings.

\section{Methods}
This section conducts detailed experiments on various deep learning models such as UNet, Res-UNet, Attention Res-UNet, and nnUNet. It explores their network architectures, layer filter characteristics, layer interconnections, and specific functionalities.
\subsection{UNet}
UNet \cite{b11} was a pioneering CNN architecture specifically designed for biomedical image segmentation. It features an encoder-decoder structure with skip connections that allow low-level details from the encoder to be fused with high-level semantic features from the decoder. The contracting encoder path consists of repeated convolution, batch normalization, and max-pooling layers to extract abstract representations. The expansive decoder path comprises transposed convolutions and upsampling layers to recover the original resolution. Skip connections concatenate encoder and decoder features at each level, preventing loss of spatial information and enabling precise boundary localization. In this project, the architecture's kernel size and filter count are set at 3 across all convolution layers, suited for image segmentation. The network begins with 64 filters in the first layer, increasing progressively in deeper layers to promote hierarchical learning. After 4 decoding layers, the output moves through a final fully connected convolutional layer, where the kernel size adjusts based on the number of classes in the mask to meet specific task needs. The output is then activated using a function that varies with the number of labels, ensuring customized results for each task. 
\subsection{Res-UNet}
Res-UNet \cite{b12} extended the UNet by incorporating residual connections into its encoder path. Each encoder block contains residual units with identity mappings, facilitating gradient flow and enabling training of deeper networks. The decoder mirrors the encoder with upsampling and concatenation of skip connections. Residual blocks aid in reusing and refining features across layers.
\subsection{Attention Res-UNet}
Attention Res-UNet \cite{b13} further enhanced the Res-UNet by integrating attention gates after each decoder stage. These compute attention maps highlighting salient regions, which are element-wise multiplied with features from skip connections. This focuses the model on relevant areas and suppresses noise, improving segmentation of fine details. The encoder follows the Res-UNet design, while the decoder incorporates attention gates and skip connections.
\subsection{nnUNet}
nnUNet  \cite{b14} is a deep learning framework designed for medical image segmentation, aiming to simplify model configuration and optimization while improving performance. nnUNet is a self-configuring framework based on the UNet architecture and robust training schemes. Compared to the original UNet model, nnUNet has undergone only minor modifications. For example, it employs instance normalization instead of batch normalization, and replaces the standard ReLU activation function with leaky ReLUs. Moreover, the framework explicitly outlines a series of steps required for model training. These steps have a significant impact on the model's performance. For instance, in the preprocessing stage, operations like resampling and regularization are included to improve data quality and model robustness. During training, aspects such as the selection of loss functions, configuration of optimizers, and application of data augmentation strategies are involved to enhance the model's learning efficacy and generalization ability.

The general settings of the UNet, Res-UNet, Attention Res-UNet, and nnUNet models in our experiments are described as follow.

For brain tumour and polyp segmentation, the preprocessing includes resizing to 256×256, standardizing, and normalizing to a 0-1 range. These images are fed into the UNet, Res-UNet, and Attention Res-UNet, all using an Adam optimizer with a 1e-5 learning rate and callbacks like Early Stopping. For cardiac segmentation, due to MRI's grayscale, output masks are (128, 128, 4) with inputs at (128, 128, 1), using a Softmax for multi-class segmentation. The nnUNet model follows identical data processing steps. As previously mentioned, nnUNet can automatically specify settings for all phases of segmentation tasks, achieving task-specific optimization without the need for manual adjustments. In this project, the default nnUNet was adopted as the primary network architecture due to its outstanding performance in numerous segmentation tasks. The nnUNet architecture consists of an encoder, a decoder and transposed convolutions. Additionally, nnUNet applies normalization to the target voxels of each dataset (image) by subtracting the mean and dividing by the standard deviation, while the non-target (background) voxels are kept at 0. Aligning with nnUNet's initial configurations, downsampling is performed using strided convolutions, and upsampling is executed through transposed convolutions. The input patch size is chosen as 128×128×128, with a batch size of 2. The process includes five downsampling stages, culminating in a bottleneck feature map size of 4×4×4. The initial count of convolutional kernels starts at 32 and doubles after each downsampling stage, reaching up to a maximum of 320. The decoder's kernel count mirrors that of the encoder. Leaky ReLUs are selected for nonlinear activation, and instance normalization is employed for normalizing the feature maps. The training extends over 1000 epochs.

\section{Brain Tumour Segmentation}
In the medical field, brain tumour segmentation is considered one of the most challenging tasks. Accurately delineating the tumour boundaries is crucial not just for diagnosis but also plays a pivotal role in developing treatment plans and monitoring treatment progress, thereby significantly enhancing the quality of patient care \cite{b26, b27, b28}. This experiment aims to explore the accuracy of delineating the shape and boundaries of lower-grade gliomas (LGGs) , which are slowly developing brain tumours that are generally less aggressive and have a better prognosis compared to higher-grade tumours, in brain MRI scans using models such as UNet, Res-UNet, Attention Res-UNet, and nnUNet.

\subsection{Dataset}
The brain tumour segmentation data, sourced from Kaggle (https://www.kaggle.com/datasets/mateuszbuda/lgg-mri-segmentation) and originally from The Cancer Imaging Archive (TCIA), features MRI brain images from 110 patients, totaling 7858 images. These images are evenly divided between brain images and their FLAIR abnormality segmentation masks. For the experiment, 1200 images and masks were randomly selected. To focus on abnormality segmentation, images and masks without low-grade gliomas abnormalities (referred to as negative) were excluded, leaving only positive image-mask pairs. The dataset was then divided into training, testing, and validation sets in an 8:1:1 ratio, resulting in training, validation, and testing cases for the UNet, Res-UNet, and Attention Res-UNet models. The nnUNet model, employing 5-fold cross-validation, drew on the combined cases from the training and validation sets of the other models, automatically partitioning them for training and validation. In all four models, the testing set is consistent.
\subsection{Result}
In this experiment, the performance metrics include the precision, recall, accuracy, DSC and IoU. The following Table I displays information related to these performance metrics. From the performance metrics for four different neural network models, deductions can be made based on the information provided in the table.\\
\begin{table}[btp]
\caption{Brain tumour segmentation: Performance metrics for UNet, Res-UNet, Attention Res-UNet and nnUNet on testing dataset.}
\begin{center}
\begin{tabular}{cccccc}
\hline
\textbf{Model}              & \textbf{Precision} & \textbf{Recall} & \textbf{Accuracy} & \textbf{DSC}& \textbf{IoU}   \\ \hline
\textbf{UNet}& \textbf{0.884}     & 0.767           & 0.990             & 0.821          & 0.697          \\
\textbf{Res-UNet}           & 0.873              & 0.853           & 0.992             & \textbf{0.863}          & \textbf{0.759}          \\
\textbf{Attention Res-UNet} & 0.842              & 0.806           & 0.989             & 0.824& 0.700          \\
\textbf{nnUNet}& 0.856              & \textbf{0.881}  & \textbf{0.994}    & 0.850& 0.757\\ \hline
\end{tabular}
\end{center}
\end{table}    

\begin{enumerate}
    \item UNet: It demonstrates high precision, effectively identifying true positives among its predictions. Yet, its recall is lower than other models, indicating it may overlook more true positives (false negatives). Additionally, both its DSC and IoU scores are the lowest, reflecting poorer overlap with the ground truth compared to other models.
    \item Res-UNet: it has the highest DSC score (0.863) and IoU (0.759), indicating the best segmentation overlap, making it a strong model for tasks where accurate delineation of tumour boundaries is crucial.
    \item Attention Res-UNet: It has the lowest precision at 0.842 and the lowest accuracy at 0.989. Despite this, an accuracy of 0.989 is still very high, indicating that the model performs well overall, albeit slightly less so than its counterparts in these particular metrics.
    \item nnUNet: It stands out with the highest recall (0.881) and accuracy (0.994), suggesting it is the most reliable at identifying true positives and overall prediction correctness. This makes it a potentially valuable model for clinical applications, where missing the presence of a tumour (false negatives) could have significant consequences.
    \item Similarity in Precision, Accuracy, and DSC Scores: Despite minor differences in precision, accuracy, and DSC scores among the models, their close performance indicates all are capable of accurately predicting true positives, identifying cases, and matching tumour boundaries well. This suggests any model could be fit for brain tumour segmentation, with selection dependent on factors like computational speed and implementation ease.
    \item Significant Variability in Recall and IoU: The models show notable differences in recall and IoU, indicating varied effectiveness in capturing all relevant tumour cases and accurately outlining tumour areas against ground truth. High recall is essential in clinical settings to avoid missing tumours, essential for treatment planning, pointing to some models (e.g., nnUNet) as more appropriate for this task. IoU variations reveal differences in segmentation precision, critical for precise treatment or surgical planning, with models like Res-UNet and nnUNet showing superior accuracy in delineating tumour boundaries.
    \item Generally Lower IoU Scores: IoU scores being not quite high across all models could be due to several factors inherent to the task of brain tumour segmentation: complexity of tumour shapes(irregular and complex shapes); variability in tumour appearance; image quality and resolution; overlap with non-tumour tissues (tumours might have similar intensity values to surrounding tissues).
    \item DSC Coefficient vs. IoU: Minor variations in DSC scores compared to IoU suggest models capture tumours well but differ more in precisely outlining tumour boundaries, as IoU indicates. This is crucial for treatment planning and monitoring tumour growth, highlighting IoU’s importance in model evaluation. In essence, despite DSC and IoU assessing overlap, IoU’s calculation highlights more significant boundary delineation differences among models, essential for clinical application insights. Improving IoU scores might involve incorporating more sophisticated image processing techniques, enhancing the models with additional contextual information, or using higher-quality imaging data. It may also benefit from more advanced training strategies that specifically target the improvement of segmentation boundary accuracy, such as utilizing more refined loss functions or incorporating post-processing techniques to refine the segmentation outputs.
\end{enumerate}
Four examples were selected to demonstrate the predictive results of the four models throughout the experiment, including the provided images and their corresponding ground truth masks (see Fig. 1).
\begin{figure}[btp]
\centering
\includegraphics[width=0.9\linewidth]{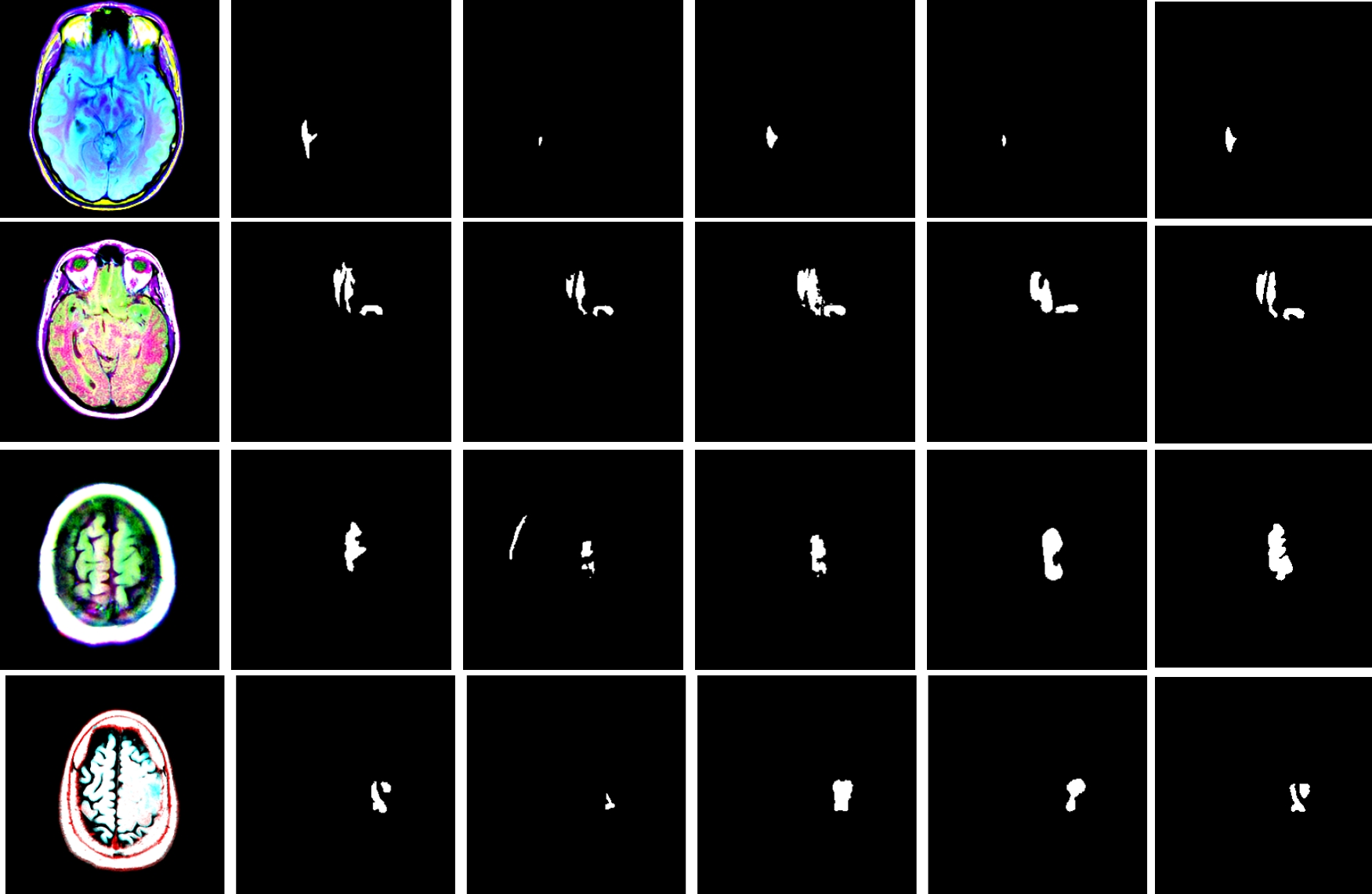}
\caption{\label {fig.1:} Brain tumour segmentation: Four examples showing segmentation by four models. From left to right: input images, ground-truth, and segmentation by UNet, Res-UNet, Attention Res-UNet, nnUNet, in order.}
\end{figure}

\section{Polyp Segmentation}
Polyps are a leading cause of colorectal cancer, with colonoscopy being the preferred detection and removal method. Accurately identifying polyps is key to diagnosing and treating colorectal cancer, yet segmenting them during colonoscopy is difficult due to their varied shapes, sizes, and color contrasts \cite{b29, b30}. This study focuses on precisely defining intestinal polyps' contours and boundaries using models like UNet, Res-UNet, Attention Res-UNet, and nnUNet.

\subsection{Dataset}
CVC-ClinicDB \cite{b31} is a dataset of frames from colonoscopy videos, containing original images and corresponding polyp masks (ground truth). It comprises 612 original images paired with 612 masks. In this experiment, the data is split into training, validation, and test sets in an 8:1:1 ratio. Similarly to the previous brain segmentation experiment, all models (UNet, Res-UNet, Attention Res-UNet, and nnUNet) use the same dataset. During nnUNet training, it receives the combined training and validation data from the other models. In all four models, the testing set is consistent.
\subsection{Result}
The following Table II displays performance metrics including precision, recall, accuracy, DSC, and IoU for four different models. Deductions can be made based on the information provided in the table.

\begin{table} [hbtp]
\caption{Polyp segmentation: Performance metrics for UNet, Res-UNet, Attention Res-UNet and nnUNet on testing dataset.}
\centering
\begin{tabular}{cccccc}
\hline
\textbf{Model}              & \textbf{Precision} & \textbf{Recall} & \textbf{Accuracy} & \textbf{DSC}& \textbf{IoU}   \\ \hline
\textbf{UNet}& 0.868              & 0.764           & 0.969             & 0.813          & 0.685          \\
\textbf{Res-UNet}           & 0.907              & 0.751           & 0.971             & 0.821          & 0.697          \\
\textbf{Attention Res-UNet} & 0.902              & 0.754           & 0.971             & 0.821          & 0.697          \\
\textbf{nnUNet}& \textbf{0.940}     & \textbf{0.949}  & \textbf{0.993}    & \textbf{0.941} & \textbf{0.895} \\ \hline
\end{tabular}
\end{table}

\begin{enumerate}
    \item Precision and Accuracy: All four models excel in precision and accuracy, although UNet has the lowest scores (0.868 precision and 0.969 accuracy, but the values are still quite high). Yet, the minor differences in these metrics among the models imply a consistent strength in identifying true positives across tested datasets.
    \item Recall, DSC, and IoU: The UNet, Res-UNet, and Attention Res-UNet models show similar recall, DSC, and IoU scores, evaluating their ability to detect all relevant cases (recall) and their segmentation accuracy against the ground truth (DSC and IoU). Despite being lower than nnUNet's scores, they remain acceptable. Nonetheless, in critical fields like medical imaging where overlooking a true positive can be crucial, even minor discrepancies matter.
    \item nnUNet Performance: The nnUNet excels, showing the highest precision (0.940), recall (0.949), accuracy (0.993), DSC (0.941) and IoU (0.895) scores, markedly surpassing other models. Its recall indicates a stronger capability in identifying actual positives, minimizing missed polyp cases. The nnUNet's DSC and IoU, significantly higher than its counterparts, suggests a closer match to the ground truth in predictions. This accuracy is critical in medical tasks for precise area segmentation.
\end{enumerate}
In summary, nnUNet outperforms in all aspects, especially in recall, DSC and IoU, crucial for accurate detection and segmentation. It emerges as the top choice for polyp segmentation, provided its computational demands are manageable. Other models, while effective, serve as alternatives under nnUNet's resource or speed constraints, or specific clinical needs.
Based on the results of the above polyp segmentation experiment, four examples were selected to demonstrate the performance of the four models in polyp segmentation. From these four examples, it is clear that nnUNet has the best performance (see Fig. 2).
\begin{figure}[btp]  
\centering
\includegraphics[width=0.9\linewidth]{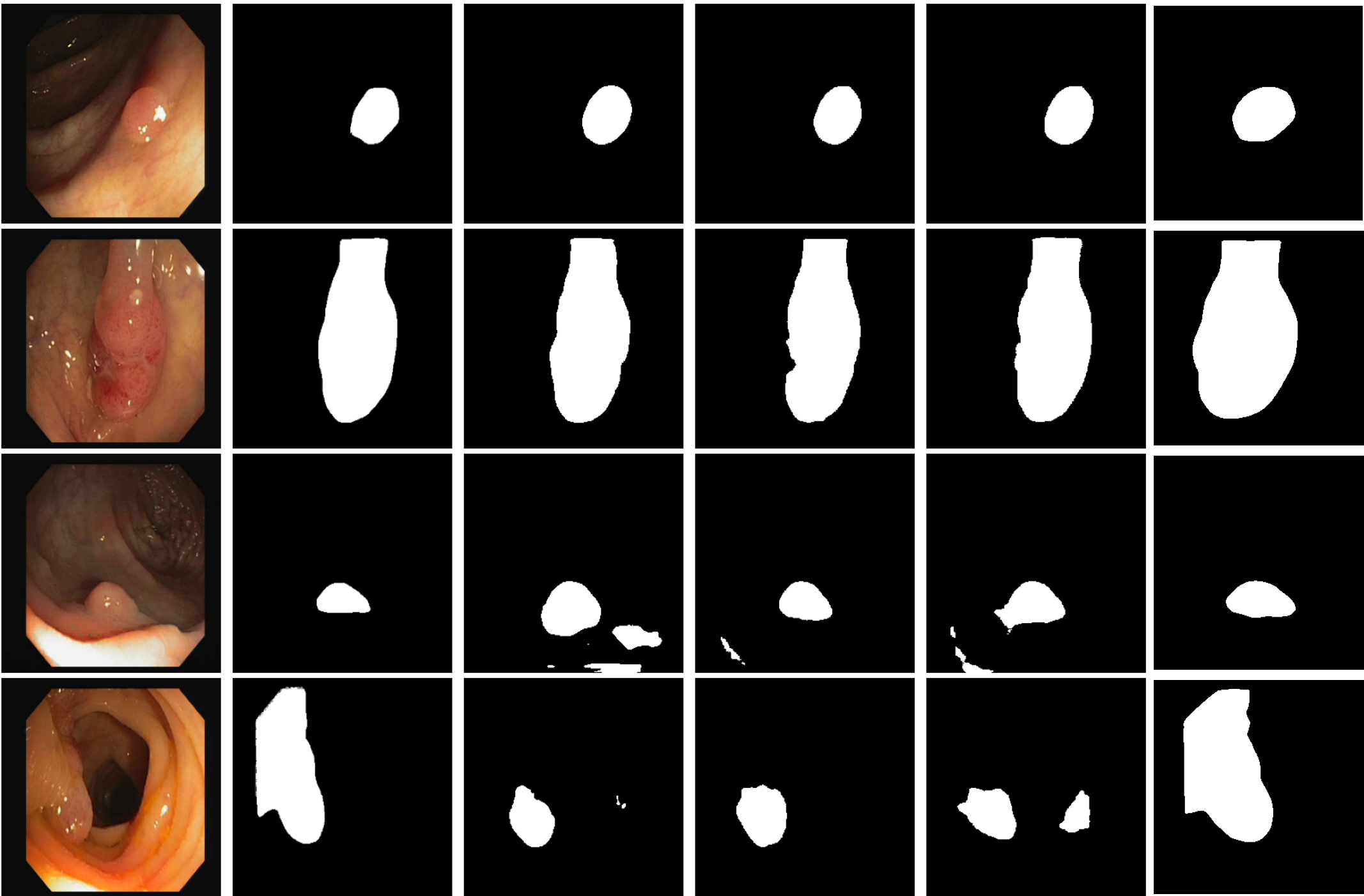}
\caption{\label{fig. 2:} Polyp segmentation: Four examples showing segmentation by four models. From left to right: input images, ground-truth, and segmentation by UNet, Res-UNet, Attention Res-UNet, nnUNet, in order.}
\end{figure}

\section{Heart Segmentation}
Medical imaging of the heart is essential for diagnosing and treating cardiovascular diseases, offering key insights into cardiac structure and function. Precise segmentation is critical for accurately measuring features like chambers, wall thickness, and blood flow dynamics, essential for diagnosing heart failure, arrhythmias, and other cardiac conditions \cite{b40},\cite{b41}. Yet, this process is challenging due to the heart's complex anatomy and motion artifacts from dynamic imaging. Manual segmentation, while precise, is time-consuming and subject to variability \cite{b33}, \cite{b34}. Addressing this, our study evaluates deep learning models—UNet, Res-UNet, Attention Res-UNet, and nnUNet—for automating heart image segmentation. 
\subsection{Dataset}
This study utilized the ACDC dataset \cite{b35} from the University Hospital of Dijon, featuring cardiac MRI of 150 patients for cardiac segmentation. It spans diastole and systole phases, with 100 and 50 patients in the training and testing sets, respectively. Sets include original images and masks for the diastole phase, categorizing heart parts from 0 to 3 for background, RV, Myo, and LV. The goal was multi-class segmentation during the diastole phase. After preprocessing, the dataset comprised 951 images and masks, divided into training, validation, and testing sets in an 8:1:1 ratio. The UNet, Res-UNet, Attention Res-UNet, and nnUNet models were trained on these sets, with nnUNet utilizing combined training and validation sets. All models were tested using a consistent set of images and masks.
\subsection{Result}
From performance metrics of precision, recall, DSC, and IoU for four different models, deductions can be made based on the information provided in the Table III:

\begin{table*} [btp]
\caption{Heart segmentation: Precision, Recall, DSC and IoU score for each class by four models.} 
\begin{tabular}{|c|cccc|cccc|cccc|cccc|}
\hline
                            & \multicolumn{4}{c|}{\textbf{Precision}}                                                                                   & \multicolumn{4}{c|}{\textbf{Recall}}                                                                                      & \multicolumn{4}{c|}{\textbf{DSC}}                                                                                           & \multicolumn{4}{c|}{\textbf{IoU}}                                                                                            \\ \hline
\textbf{Model}              & \multicolumn{1}{c|}{\textbf{0}} & \multicolumn{1}{c|}{\textbf{1}}    & \multicolumn{1}{c|}{\textbf{2}}    & \textbf{3}    & \multicolumn{1}{c|}{\textbf{0}} & \multicolumn{1}{c|}{\textbf{1}}    & \multicolumn{1}{c|}{\textbf{2}}    & \textbf{3}    & \multicolumn{1}{c|}{\textbf{0}} & \multicolumn{1}{c|}{\textbf{1}}     & \multicolumn{1}{c|}{\textbf{2}}     & \textbf{3}     & \multicolumn{1}{c|}{\textbf{0}} & \multicolumn{1}{c|}{\textbf{1}}     & \multicolumn{1}{c|}{\textbf{2}}     & \textbf{3}     \\ \hline
\textbf{UNet}& \multicolumn{1}{c|}{0.99}       & \multicolumn{1}{c|}{0.93}          & \multicolumn{1}{c|}{0.87}          & 0.97          & \multicolumn{1}{c|}{0.99}       & \multicolumn{1}{c|}{\textbf{0.92}} & \multicolumn{1}{c|}{0.86}          & 0.94          & \multicolumn{1}{c|}{0.993}      & \multicolumn{1}{c|}{0.924}          & \multicolumn{1}{c|}{0.869}          & 0.956          & \multicolumn{1}{c|}{0.987}      & \multicolumn{1}{c|}{0.860}          & \multicolumn{1}{c|}{0.768}          & 0.916          \\ \hline
\textbf{Res-UNet}           & \multicolumn{1}{c|}{0.99}       & \multicolumn{1}{c|}{\textbf{0.95}} & \multicolumn{1}{c|}{\textbf{0.89}} & \textbf{0.98} & \multicolumn{1}{c|}{1.00}       & \multicolumn{1}{c|}{\textbf{0.92}} & \multicolumn{1}{c|}{0.85}          & 0.94          & \multicolumn{1}{c|}{0.994}      & \multicolumn{1}{c|}{\textbf{0.936}} & \multicolumn{1}{c|}{0.869}          & \textbf{0.961} & \multicolumn{1}{c|}{0.988}      & \multicolumn{1}{c|}{\textbf{0.879}} & \multicolumn{1}{c|}{0.769}          & \textbf{0.924} \\ \hline
\textbf{Attention Res-UNet} & \multicolumn{1}{c|}{0.99}       & \multicolumn{1}{c|}{0.94}          & \multicolumn{1}{c|}{\textbf{0.89}} & 0.96          & \multicolumn{1}{c|}{0.99}       & \multicolumn{1}{c|}{\textbf{0.92}} & \multicolumn{1}{c|}{0.86}          & \textbf{0.96} & \multicolumn{1}{c|}{0.994}      & \multicolumn{1}{c|}{0.930}          & \multicolumn{1}{c|}{0.874}          & 0.960          & \multicolumn{1}{c|}{0.988}      & \multicolumn{1}{c|}{0.869}          & \multicolumn{1}{c|}{0.777}          & \textbf{0.924} \\ \hline
\textbf{nnUNet}& \multicolumn{1}{c|}{1.00}          & \multicolumn{1}{c|}{0.82}          & \multicolumn{1}{c|}{\textbf{0.89}} & 0.95          & \multicolumn{1}{c|}{1.00}        & \multicolumn{1}{c|}{0.81}          & \multicolumn{1}{c|}{\textbf{0.90}} & 0.95          & \multicolumn{1}{c|}{0.996}      & \multicolumn{1}{c|}{0.812}          & \multicolumn{1}{c|}{\textbf{0.894}} & 0.953          & \multicolumn{1}{c|}{0.992}      & \multicolumn{1}{c|}{0.776}          & \multicolumn{1}{c|}{\textbf{0.818}} & 0.921          \\ \hline
\end{tabular}
\end{table*}

\begin{enumerate}
    \item Background (Class 0): Not surprisingly, all models excel at identifying the background class due to its larger area and simpler texture, making it easier to learn and segment compared to anatomical structures.
    \item Right Ventricle (RV, Class 1): The nnUNet's lower scores in precision, recall, DSC, and IoU for RV segmentation suggest it is less suitable for this task compared to others. Res-UNet and Attention Res-UNet both excel in heart segmentation, with Res-UNet slightly outperforming Attention Res-UNet in certain metrics.
    \item Myocardium (Myo, Class 2): Segmenting the Myocardium (Myo) proves more difficult for the models, showing lower metrics than the background, RV, and LV classes, likely due to Myo's smaller size and complex shape. However, the nnUNet stands out for its superior Myo segmentation, indicating its adaptive architecture and training strategy successfully tackle Myo's unique features.
    \item Left Ventricle (LV, Class 3): Res-UNet and Attention Res-UNet indeed have similar, high performance metrics for LV segmentation. UNet and nnUNet also perform well, with only slight differences compared to the other two models. This implies that the LV, while segmented reasonably well by all models, benefits from the features of the Res-UNet and Attention Res-UNet architectures. 
    \end{enumerate}
Based on the results of the above heart segmentation experiment, four examples were selected to demonstrate the performance of the four models in heart segmentation (see Fig. 3).
\begin{figure}[btp]
\centering
\includegraphics[width=0.9\linewidth]{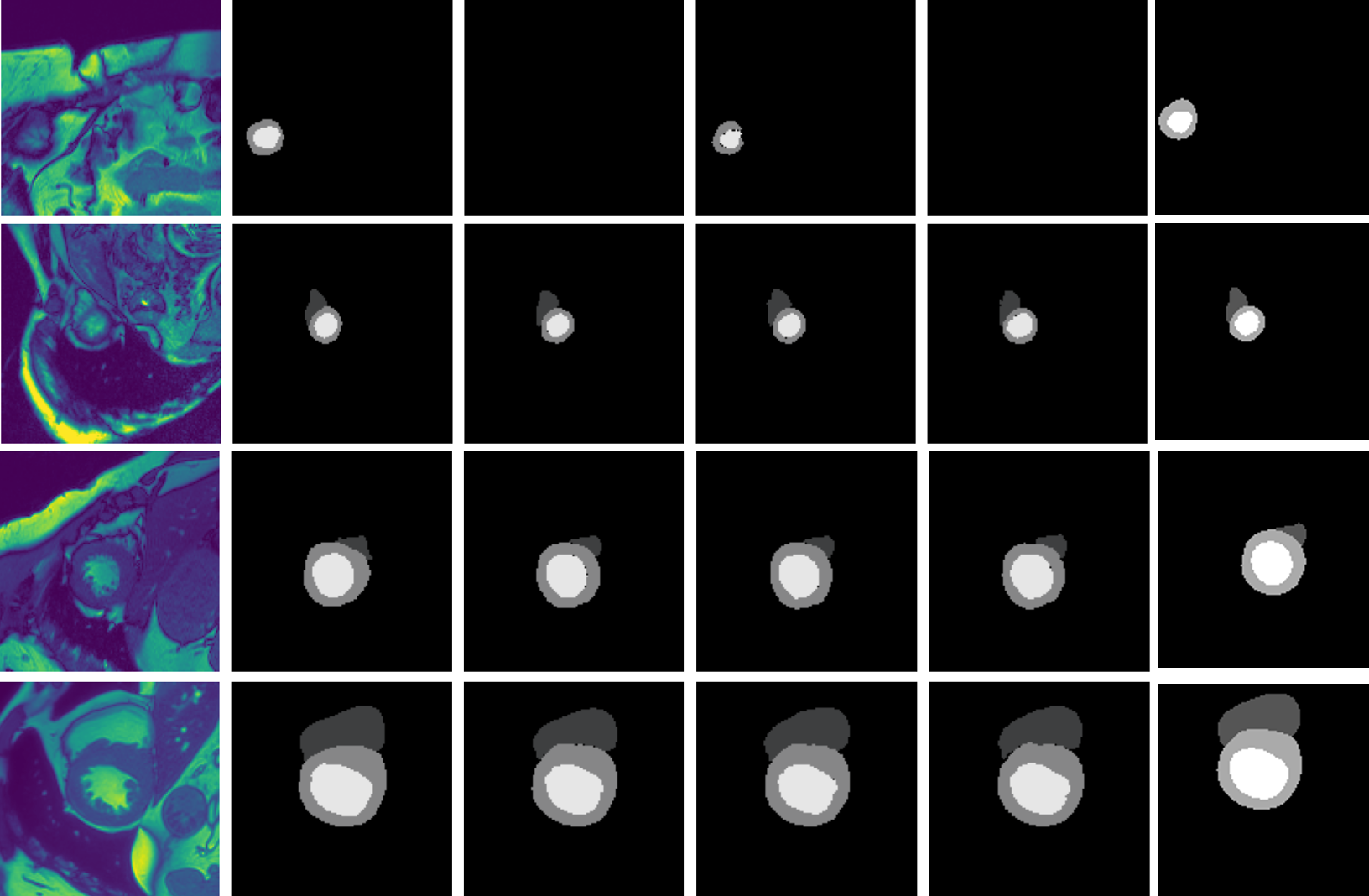}
\caption{\label{fig. 3}Heart segmentation: Four examples showing segmentation by four models. From left to right: input images, ground-truth, and segmentation by UNet, Res-UNet, Attention Res-UNet, nnUNet, in order.}
\end{figure}

\section{Discussion }
This study provides a thorough comparison of four deep learning models—UNet, Res-UNet, Attention Res-UNet, and nnUNet—on three medical imaging tasks: brain tumour, polyp, and multi-class heart segmentation.
\begin{enumerate}
    \item Brain Tumour Segmentation: In brain tumour segmentation, Res-UNet and nnUNet excelled, showing superior performance in critical metrics. Res-UNet achieved the highest DSC and IoU scores, indicating exceptional accuracy in tumour boundary delineation. nnUNet was notable for its high recall and accuracy, proving to be highly reliable in identifying tumour presence, which is vital for effective clinical diagnosis and treatment planning. These models’ advanced architectures allow for superior capture of complex image details, crucial for accurate medical image analysis. Despite nnUNet’s intensive computational demands, its outstanding overall performance highlights its significant potential in medical applications. 
    \item Polyp Segmentation: The nnUNet model outperformed others, exhibiting exemplary precision and recall, indicating its potential as a reliable tool for endoscopic analysis, provided the computational resources for its extensive training are available.
    \item Heart Multi-Class Segmentation: In the segmentation of cardiac structures, Res-UNet and Attention Res-UNet showed promising results, especially in LV segmentation, where precision in anatomical delineation is paramount. As for segmenting the RV, Res-UNet stands out as the best performer in this class, suggesting that the residual connections in its architecture are beneficial for capturing the complex features of the RV. The nnUNet’s performance, particularly in segmenting the myocardium with high DSC and IoU scores, suggests its utility in scenarios where the most detailed tissue differentiation is required.
\end{enumerate}
\section{Conclusions}
The discussion above separately addresses brain segmentation, polyp segmentation, and heart multi-class segmentation. From the three experiments, it is evident that Res-UNet performs better than UNet and Attention Res-UNet, achieving higher scores in certain performance metrics than nnUNet. Nevertheless, when combining the results from all three experiments, nnUNet emerges as the best model for the following reasons:
\begin{enumerate}
    \item Although nnUNet requires considerable computational resources, in clinical applications, high diagnostic accuracy and low misdiagnosis rates are paramount. For example, missing a tumour in a patient can lead to delayed treatment with serious consequences.
    \item In brain segmentation, the DSC score (0.863) and IoU (0.759) for Res-UNet are higher than those for nnUNet (0.850 and 0.757, respectively), though the margins are very small. However, the values of recall and accuracy for nnUNet are higher than those for Res-UNet.
    \item In polyp detection, all performance metrics for nnUNet are higher than the other three models, especially recall, DSC, and IoU, where nnUNet's scores (0.949, 0.941, and 0.895, respectively) far surpass those of the others.
    \item In heart multi-class segmentation, nnUNet's metrics for RV are the lowest at 0.82, 0.81, 0.812, and 0.776, but still high and acceptable. nnUNet leads in myocardial segmentation with the highest scores, including a notable recall of 0.90 and an IoU of 0.818. In LV segmentation, while Res-UNet and Attention Res-UNet perform slightly better, nnUNet's results are closely comparable, nearly matching those models.
\end{enumerate}
This study sets a  benchmark in the field of medical image segmentation and offers  insights to future researchers on key considerations when applying UNet and its variants to medical image analysis. By systematically comparing models such as UNet, Res-UNet, Attention Res-UNet, and nnUNet, the research highlights their respective strengths and potential applications in segmenting MRI and endoscopic images, as well as in binary and multi-class tasks.

However, despite notable progress in two-dimensional image segmentation, there are limitations in the types of data handled. In real medical scenarios, many critical datasets typically exist in three-dimensional forms, yet all experiments conducted in this study are limited to two-dimensional data. This limitation could restrict the models' ability to address real medical imaging challenges effectively. Future research could also explore more extended models similar to nnUNet, and develop or fine-tune large AI models, to enhance their capability to process and analyze three-dimensional medical images.
Additionally, this experiment faces a data leakage issue due to the division of data at the image level rather than the patient level. To prevent such problems in future work, it is recommended: 1. Strict data segmentation ensures that all images from the same patient are contained within a single dataset (training, validation, or test) before training begins. 2. Proper data handling before augmentation to ensure that enhancements are only applied to the training set, thus avoiding indirect data leaks into validation or test sets. 

\bibliographystyle{ieeetr}
\bibliography{Ref}

\end{document}